\documentclass[aps,prl,twocolumn,epsfig,preprintnumbers,superscriptaddress,10pt]{revtex4}

\usepackage{graphicx}
\usepackage{dcolumn}
\usepackage{bm}

\newcommand{\be}{\begin{equation}}
\newcommand{\ee}{\end{equation}}
\newcommand{\bea}{\begin{eqnarray}}
\newcommand{\eea}{\end{eqnarray}}
\newcommand{\ba}{\begin{eqnarray}}
\newcommand{\ea}{\end{eqnarray}}

\newcommand{\beq}{\begin{equation}}
\newcommand{\eeq}{\end{equation}}
\newcommand{\beqa}{\begin{eqnarray}}
\newcommand{\eeqa}{\end{eqnarray}}
\newcommand{\beqar}{\begin{eqnarray*}}
\newcommand{\eeqar}{\end{eqnarray*}}

\begin{document}

\preprint{CERN-PH-TH/2010-288}

\title{Jet Quenching via Jet Collimation}

\author{Jorge Casalderrey-Solana} 
\affiliation{    Physics Department, 
    Theory Unit, CERN,
    CH-1211 Gen\`eve 23, Switzerland}
    
\author{Jos\'e Guilherme Milhano}

\affiliation{    Physics Department, 
    Theory Unit, CERN,
    CH-1211 Gen\`eve 23, Switzerland}

\affiliation{ CENTRA, Departamento de F\'isica, Instituto Superior T\'ecnico (IST),
Av. Rovisco Pais 1, P-1049-001 Lisboa, Portugal}

\author{Urs Achim Wiedemann}
\affiliation{    Physics Department, 
    Theory Unit, CERN,
    CH-1211 Gen\`eve 23, Switzerland}


\begin{abstract}
The ATLAS Collaboration recently reported strong modifications of dijet properties in
heavy ion collisions. In this work, we discuss to what extent these first data constrain already the microscopic
mechanism underlying jet quenching. Simple kinematic arguments lead us to identify a frequency
collimation mechanism via which the medium efficiently trims away the soft components of the jet
parton shower. Through this mechanism, the observed dijet asymmetry can be accomodated 
with values of $\hat{q}\, L$ that lie in the expected order of magnitude. 
\end{abstract}

\maketitle

{\bf 1. Introduction.}
 `Jet quenching' refers most generally to a picture in which high-$p_T$ partons produced in hard processes
lose, in addition to their vacuum fragmentation, a significant amount of energy in interactions with the 
QCD matter produced in heavy ion collisions.  Models 
implementing such medium-induced parton energy loss can account for characteristic features of 
the data measured at the Relativistic Heavy Ion Collider (RHIC) over the last decade. 
In particular, the assumption of a significant medium-modification of the final state parton shower
is supported by the strong suppression of all high-$p_T$ hadron spectra by a factor $\sim 5$ in 
central heavy ion collisions compared to p-p collisions; by the absence of a comparably strong medium 
effect in the photon spectra; and by the dependence of the suppression pattern on the centrality of the 
collision.
For a more complete account of RHIC
data, and for an overview of open issues in theory and modeling of jet quenching, we refer to several
recent reviews \cite{Wiedemann:2009sh,CasalderreySolana:2007zz,d'Enterria:2009am,Majumder:2010qh,Jacobs:2004qv,Gyulassy:2003mc}. 

Almost all the experimental support for jet quenching
at RHIC comes from the study of leading single inclusive hadron spectra and high-$p_T$
triggered leading hadron correlation functions. However, to test and further constrain the basic
view of jet quenching as a medium-modified final state parton shower, it is not
sufficient to characterize the energy degradation of the leading fragment in the shower.
One needs also to characterize  the pattern of energy redistribution amongst fragments 
without experimental biases on the
fragmentation pattern. That is to say by identifying jet fragmentation patterns in heavy ion collisions
not via trigger particles but via calorimetric jet measurements. Motivated by these considerations,
first calorimetric  jet measurements at
RHIC have become available in the last two years \cite{Ploskon:2009zd,Bruna:2009em,Lai:2009zq}.
However, the maximal center-of-mass energy $\sqrt{s_{\rm NN}} = 200$ GeV  in
Au-Au collisions at RHIC has limited, so far, these studies to
jet energies $E_T < 40$ GeV. Moreover, the small cross sections for such jets at RHIC clearly
limit the possibility to perform detailed studies.

Only a few weeks after the start of the LHC heavy ion programme, the ATLAS collaboration released
first data on the medium-induced modification of calorimetrically measured jets
in Pb-Pb collisions at $\sqrt{s}=2.76$ TeV~\cite{Atlas:2010bu} (see also preliminary data
from the CMS collaboration~\cite{Bolek}).
In these heavy ion collisions at the TeV scale, one finds already in the first  $1.7\, \mu b^{-1}$ of data
abundant samples of jets with $E_T > 100$ GeV, supporting the expectation that jet samples with
$E_T > 300$ GeV are finally in experimental reach for runs at design luminosity. Most importantly, the
first ATLAS data indicate that for jets of $\mathcal{O}(100)$ GeV, the modifications of jet fragmentation in heavy ion
collisions are much more significant than the experimental uncertainties that may arise from identifying
jets in a high-multiplicity environment. This is the very basis for a qualitatively novel access to understanding
the microscopic dynamics of parton propagation in dense QCD matter. 

In the present rapid communication, we present a qualitative discussion of how the first ATLAS data already 
can contribute to constrain our dynamical picture of jet quenching. We start with a short discussion of the data.

{\bf 2. Discussion of the ATLAS measurements on jet quenching.}
ATLAS analyzed event samples from Pb-Pb collisions at $\sqrt{s_{NN}}=2.76$ TeV in which the transverse energy 
of the most energetic jet (leading jet) was $E_{T_1}>100 $ GeV and where a recoiling jet with transverse energy 
$E_{T_2}>E_{T, \, min}=25$ GeV could be found at azimuthal separation $\Delta \phi=\left| \phi_1-\phi_2 \right|>\pi/2$.
Both jet energies are identified with the anti-$k_T$ algorithm \cite{Cacciari:2008gp} with radius $R= 0.4$. 
Since the shape of jets reconstructed by this algorithm tends to be close to circular in the $\eta$-$\phi$-plane, 
we shall loosely refer to the region assigned to a jet as a cone of radius $R$. 

\begin{figure}
\centering
\includegraphics[angle=0,width=1.09\linewidth]{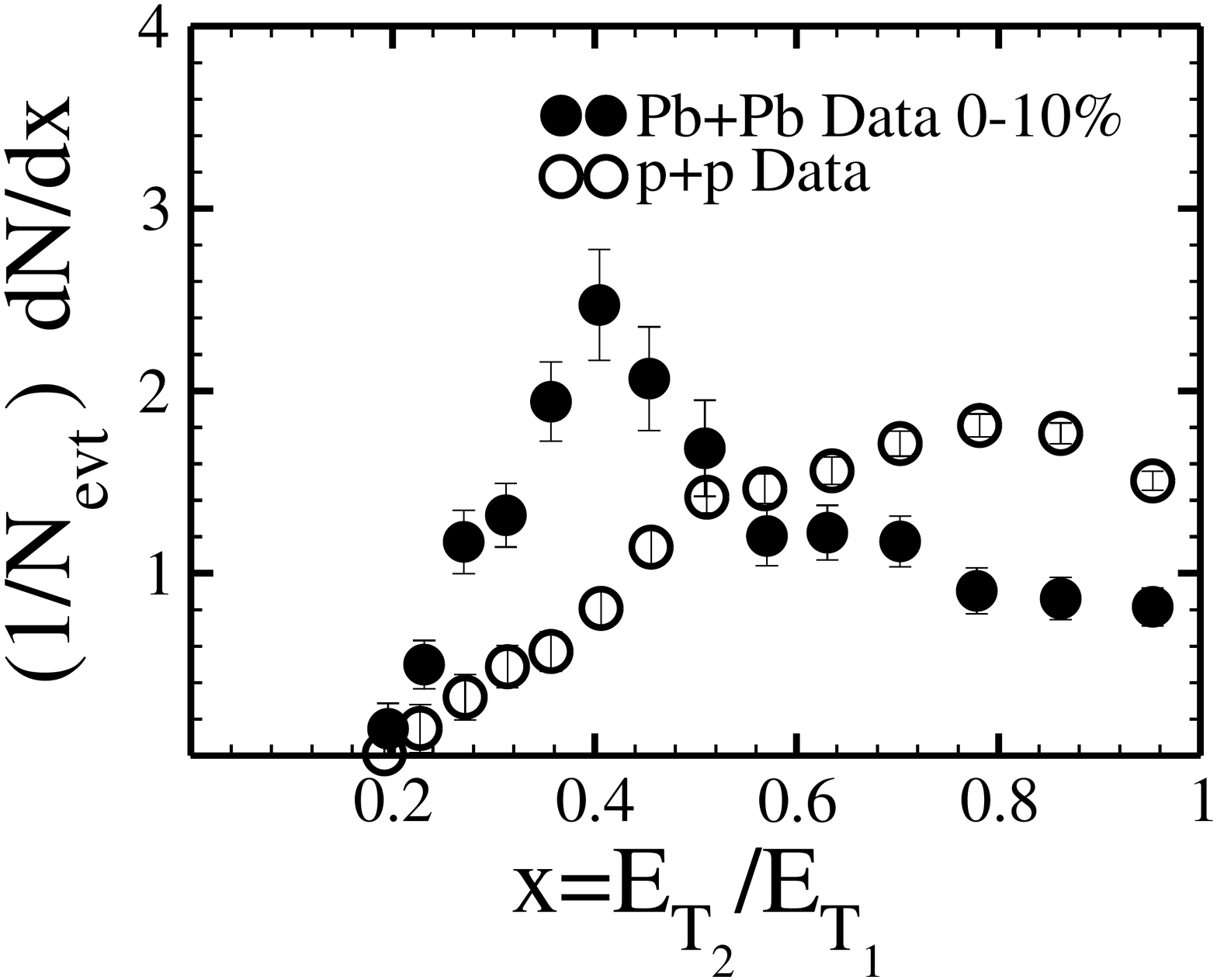}
\includegraphics[angle=0,width=0.89\linewidth]{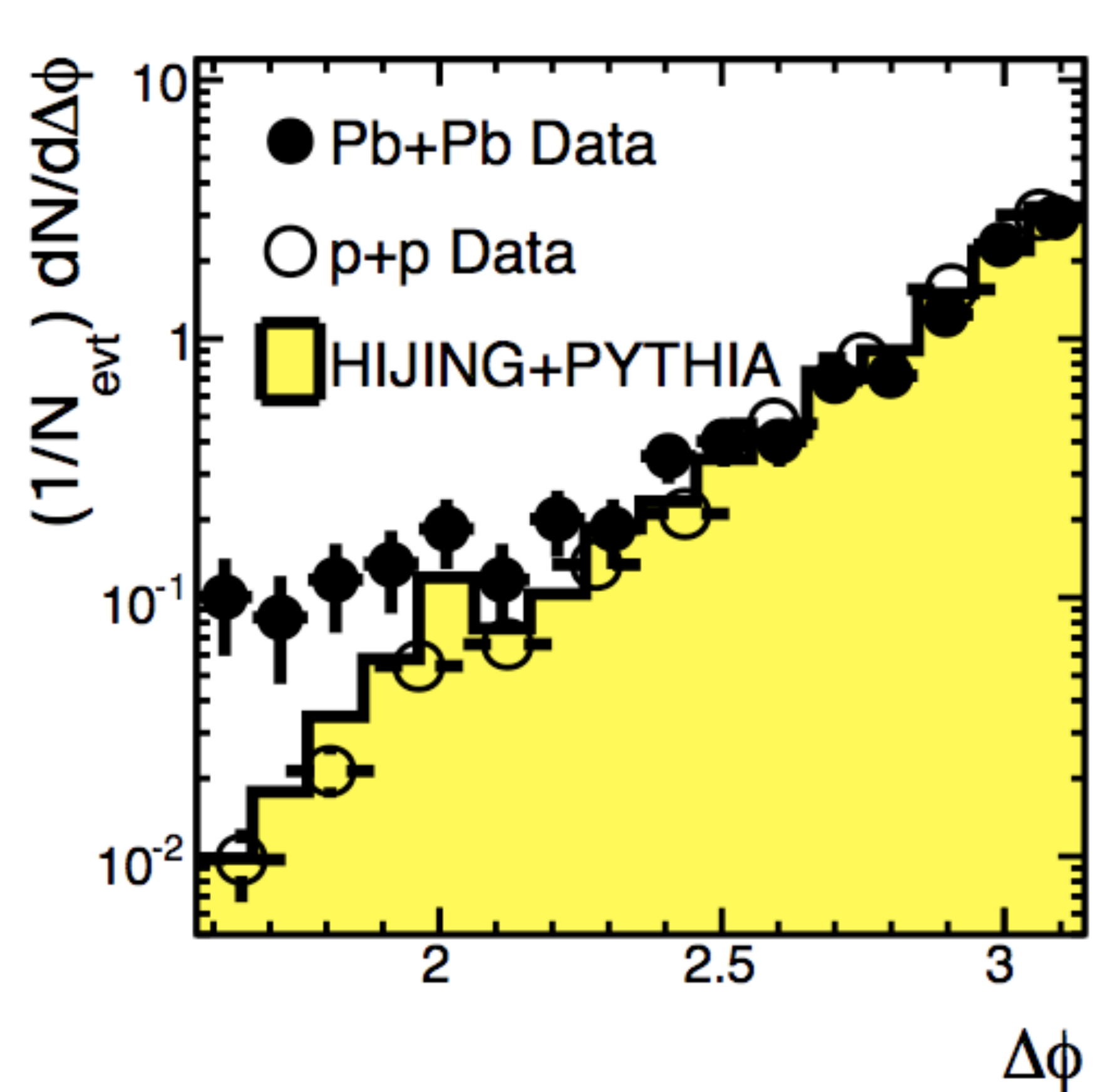}
\caption{
Top: Dijet distributions measured by ATLAS as described in the text, and plotted as a function of 
$x = E_{T_2}/E_{T_1}$ for $E_{T_1}> 100$ GeV. Data from proton-proton collisions at $\sqrt{s}=7$ TeV 
are compared to data from the 10 \% most central Pb+Pb collisions at $\sqrt{s_{\rm NN}}=2.76$ TeV. 
Bottom: The same dijet sample, plotted as a function of the azimuthal angle $\Delta \phi$
between leading and recoiling jet.}\label{fig1}
\end{figure}

In Fig.~\ref{fig1} (top) we have replotted the measured distribution of such events for the 10 \% most central 
heavy ion collisions as a function of the fraction $x$ of the leading jet energy carried by the recoiling jet, 
$x \equiv E_{T_2}/E_{T_1}$. In comparison to p-p collisions, ATLAS observes in Pb-Pb collisions 
a strong reduction of the number of recoiling jets carrying a large fraction ($x > 0.6$, say) of the maximally 
available jet energy within a cone of radius $R=0.4$, see  Fig.~\ref{fig1}(top). In contrast, the number of 
recoiling jets carrying a smaller energy fraction ($x< 0.5$, say) is enhanced in Pb-Pb collisions. 
ATLAS also  shows data on the angular dijet distribution in $\Delta \phi$, see Fig.~\ref{fig1}(bottom).
This distribution displays a significant broadening in the range $\pi/2 < \Delta \phi < 3\pi/4$ that contains
a very small fraction $\mathcal{O}$(5\%) of all dijets. In the region $3\pi/4 < \Delta \phi < \pi$, where more 
than 95 \% of all dijets lie, the distribution $dN/d\Delta \phi$ broadens at best rather mildly. 

Jet finding in the high multiplicity environment of a heavy ion collision poses challenges beyond those present in p-p collisions.
This is the case since in the 10\% most central Pb-Pb collisions there are of the order of 1500 \cite{Aamodt:2010pb} charged particles with typical transverse momentum larger than 500 MeV which result in $\mathcal{O}$(100 GeV) background energy in any cone of radius $R=0.4$.
While this background energy can be subtracted, background fluctuations could lead to large artificial asymmetries in the reconstructed dijet energy.
However, the ATLAS and CMS collaborations have put such concerns largely to rest by embedding
simulated dijets without medium-modifications both into a Monte Carlo model of the background~\cite{Atlas:2010bu}
and into real event samples~\cite{Bolek}. This provides strong support to the crucial assumption that the measured asymmetries are due to medium modifications of the jet properties.

The data shown by ATLAS are normalized to the 'number of events' and not to the total number 
$N_{\rm tot}$ of leading jets with energy $E_{T_1}> 100$ GeV. For the study of jet quenching, one would 
prefer that yields are normalized to $N_{\rm tot}$. To the extent to which there are leading jets,
for which no recoiling jet with $E_{T_2}>E_{T, \, min}$ is found by the jet finding algorithm, 
$N_{\rm evt}$ will be smaller than $N_{\rm tot}$. 
Any medium-induced softening of the leading jet to values $E_{T_2} < E_{T, \, min}$ 
(but also any multiplicity-dependent inefficiency in finding jet recoils of low $E_{T_2} > E_{T, \, min}$ GeV) 
will result in a reduction of $N_{\rm evt}$ that is more pronounced in Pb-Pb than in p-p. 

These and other considerations prompt us to limit the following discussion largely to qualitative statements. 
We start by shortly recalling the qualitative picture on which models of jet quenching
are based, before interpreting the ATLAS data in the context of this picture.

\begin{figure}
\centering
\includegraphics[angle=0,width=0.99\linewidth]{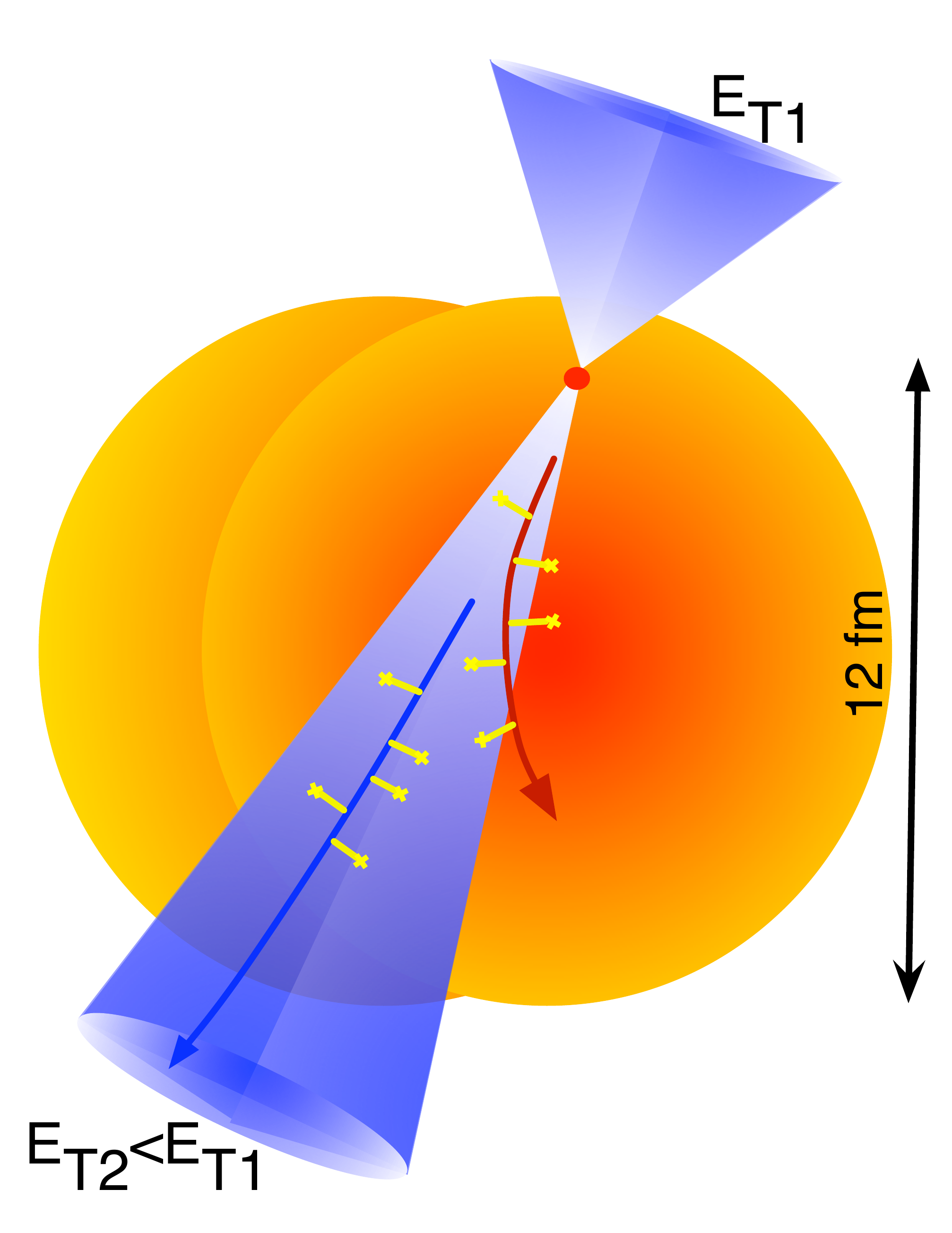}
\caption{Schematic illustration of the spatial embedding of a dijet event in a heavy ion collision.
In a central Pb+Pb collision, the overlap of the lead ions in the plane transverse to the beam direction
fills a region of more than 10 fm diameter with dense QCD matter. The leading jet and its recoil 
propagate through this matter on the way to the detectors. Compared to typical time scales in QCD,
$\mathcal{O}$(10 fm/c) is a very long time scale for interactions between a set of partonic projectiles and the 
surrounding QCD matter. This allows for strong medium-modifications of jets in heavy ion collisions.
\label{fig2}
}
\end{figure}

 {\bf 3. A qualitative picture of dijet production in heavy ion collisions.}
 Fig.~\ref{fig2} illustrates how a dijet is embedded in a heavy ion collision. 
 In the plane transverse to the beam direction, a central Pb-Pb collision
 fills a region of diameter $10$ fm with dense QCD matter. In contrast, the primary partonic process of a 
 dijet event in this region occurs at a large momentum transfer of $\mathcal{O}(E_{T})$ and is therefore localized on
 a point-like scale $\sim 1/E_{T}$ within the QCD matter. This sharp localization, illustrated
 by the red dot in Fig.~\ref{fig2},  implies  that typical soft momentum components of the surrounding QCD matter 
 cannot resolve the primary hard partonic interaction and therefore will not modify it. However, the partons 
 produced in the primary hard process may traverse a significant path length within the QCD matter, and it is
 during this final state propagation that the medium can modify the jet structure. 
 
 We note that already in proton-proton
 collisions, there are characteristic differences between the leading jet and its recoil. In particular, requiring a maximal
 jet energy $E_{T_1}$ within a cone of $R=0.4$, one selects jet fragmentation patterns that deposit more than
 the average jet energy fraction inside the subcone of size $R= 0.4$. 
 In the presence of medium effects, further
 trigger biases can occur. In particular, if there is a medium-induced mechanism that degrades the jet energy 
 fraction in a sub-cone as a function of in-medium path length then the leading jet will be oriented preferably 
 in a direction in which its path length is minimal. This results in a {\it surface bias} of the location of dijet events
 entering the experimental data sample. On average, the recoiling jet will see a larger in-medium path length
 and will hence suffer a more significant medium modification than the leading jet. On the other hand, there may
 also exist a significant contribution of dijet events produced tangentially to the nuclear overlap region, for which
 the in-medium path length of the recoiling jet is comparable to that of the leading jet ({\it corona effect}). A quantitative
 understanding of the medium modification of dijet events must thus rely on a detailed geometric modeling of 
 the nuclear overlap region and its strong dynamical evolution. 
  
In the context of this general picture, we emphasize some qualitative features in the data: 
That Fig.~\ref{fig1} (top) indicates a significant medium modification of recoiling jets supports the argument that 
there is a significant surface bias effect. Therefore, we expect that the structure of the triggered jets in Pb-Pb 
collision is comparable to that of triggered jets in p-p collisions (this expectation is, of course, testable experimentally). 
The data of Fig.~\ref{fig1} (top) imply then that a significant fraction of the recoil energy typically contained in a cone 
of radius $R=0.4$ has been transported out of it. And Fig.~\ref{fig1} (bottom) implies that this transverse redistribution of jet 
 energy has been accompanied, for most of the dijets, by a relatively mild change in the azimuthal correlation of the 
 center of the recoiling jet. 
 Therefore, any dynamical modeling of the ATLAS data must account for a significant medium-induced transverse 
broadening of the multiplicity distribution inside the recoiling parton shower as to displace an additional fraction 
of the jet energy away from the recoiling jet cone. But this broadening of the multiplicity distribution must be 
accompanied by a rather mild medium-induced transverse broadening of the distribution of recoiling jet
centers.

 We now turn to estimates of how much energy of the recoiling jet has been carried out of the jet cone,
 and to what extent the first ATLAS data constrain the dynamical mechanism underlying this redistribution.

{\bf 4. Some lessons from data}.
We are interested in estimating by how much the fraction of the energy radiated
outside the cone of the recoiling jet increases from p-p to central Pb-Pb collisions. We start by calculating the mean value ${\bar x}$ of the dijet distributions in Fig.~\ref{fig1} (top),
\be
\bar x =\frac{1}{N_{evt}}\int dx\, x \, \frac{dN}{dx} \, .
\ee
In general, the mean ${\bar x}$ will differ from the average fractional energy $\left<x\right> \equiv
\Delta E(R)/E_{T_1}$ radiated outside the cone of the recoiling jets. Differences between ${\bar x}$
and $\left<x\right>$ arise both because  the distribution of Fig.~\ref{fig1} is not normalized by the
number of leading jets, and it does not include recoiling jets with energy $E_{T_2} < E_{T,min}$.
Both these effects imply that  $\left<x\right>  <  \bar x$. Therefore, we use the calculated mean
as a lower bound
\be
\left< x\right>_{pp} \leq 0.67\, , \quad \left< x\right>_{PbPb} \leq 0.54\, .
\ee
The difference between ${\bar x}$ and  $\left<x\right>$ discussed above is more pronounced
in Pb-Pb than in p-p collisions. Therefore, the discrepancy between $\left< x\right>_{pp}$ and
${\bar x}_{pp}$ should be smaller than that between the values extracted for central Pb-Pb collisions, and $\left< x\right>_{pp} - \left< x\right>_{PbPb} \geq 0.13$.

Here, the average fractional energy $\left<x\right>$ was normalized to the
energy $E_{T_1}$ of the leading jet, reconstructed with a fixed cone $R=0.4$. This
is generally smaller than the total jet energy,   $E_{T} \geq E_{T_1}$.
However, as already explained, leading jets tend to maximize the energy within the cone,
which makes $E_{T_1}$ a good proxy for $E_{T}$. 
From jet shape measurements~\cite{Acosta:2005ix}
we estimate that for $E_{T_1} > 100$ GeV  we have
$1 > E_{T_1} /E_{T} > 0.8$, and 
\be
\frac{\left<\Delta E\right>}{E_T} > 0.8 \left( \left< x\right>_{pp} - \left< x\right>_{PbPb} \right) > 0.1\, .
\label{eq:lowerbound}
\ee
So, for total energies $E_T$ in excess of 100 GeV, medium effects in Pb-Pb collisions deposit at least 10 GeV 
additional energy outside the cone R=0.4 of the recoiling jet.  

The estimate Eq.~(\ref{eq:lowerbound}) is only a lower bound on the energy lost by out-of-cone radiation in the 
medium since we are assuming that all jets interact with the medium.
But as discussed above, there is the corona effect: due to geometry, a significant fraction of dijet events are produced 
tangentially to the collision region and will differ only mildly from dijets in p-p.
To give an upper bound on $\left<\Delta E\right> / E_T$, we take account of the corona effect by
an all-or-nothing mechanism, according to which only a fraction $(1-\alpha)$ of recoil jets interacts with the medium 
while the rest remain unchanged. 
From the ratios of the yields at $x\sim1$ in Fig.~\ref{fig1} we estimate that the fraction of unmodified jets 
$\alpha$ is $\alpha \lesssim 0.5$. This implies a mean loss of energy in the medium 
\be
\frac{\left<\Delta E\right>_m}{E_T} < \frac{\left<x\right>_{pp}-\left<x\right>_{PbPb}}{1-\alpha}<0.2\, .
\label{eq:upperbound}
\ee

We conclude from the above that the interaction of the jet with the medium must be able to transport at least 20 GeV 
outside the jet cone. Therefore, while the  suppression of single particle spectra at the LHC~\cite{alice}
 requires a strong degradation of 
the {\it longitudinal} distribution of jet fragments, the  ATLAS data imply a strong medium induced 
{\it transverse} broadening. 

A mechanism that leads both to large energy degradation and significant broadening is the medium induced radiation of a 
single hard parton at a large angle.
However, since a significant fraction of the jets lose more than $20 \%$ of their energy, such a mechanism would imply an azimuthal displacement of those jets by  $\Delta \phi > 0.1$ leading to a clearly distinguishable modification of the azimuthal distribution with respect to p-p. This is  in clear contradiction to the data in Fig.~\ref{fig1}. 
Thus, the constraints imposed by these data are indicative of a mechanism of energy loss involving the out of cone emission of many soft partons.

{\bf 5. The medium as a frequency collimator.} The picture that energy loss is carried by soft components
is also supported by simple kinematic arguments.
In the dense QCD matter produced in heavy ion collisions,
partonic jet fragments interact within a path length of several fm
with several constituents of the medium. We expect that these interactions show the main qualitative features of partonic
scattering cross sections, namely they are dominated by small-angle scattering and they have a relatively weak 
dependence on center of mass energy. As a consequence of the dominance of multiple small angle scattering, all 
components formed in the parton shower undergo Brownian motion. Let us denote by $\hat q$ the average
squared transverse momentum
that a partonic component accumulates per unit path-length by this Brownian motion.
In a medium of length $L$, all jet fragments will accumulate on average a transverse momentum $\sqrt{\hat q\, L}$.
Therefore, the softest jet fragments of energy 
\be
\omega \leq \sqrt{\hat q L}
\ee
will be decorrelated completely from the initial jet direction. In other words, the medium acts as an efficient {\it frequency collimator} which trims away the soft components of the jet.

The picture advocated above and sketched in Fig.~\ref{fig2} is a classical one: all partons in the parton shower 
accumulate on average the same transverse momentum due to multiple scattering in the medium. The less energetic
these components are, the more the medium-induced scattering will decorrelate them from the jet axis. 
To make this picture quantum mechanically consistent, one requires an argument that the softest partonic components 
in the parton shower can be regarded as independent quanta while propagating through the medium; that means,
soft components must decohere from the parton shower on a sufficiently short  time scale. 
This point can be argued on the basis of several QCD-based calculations of parton energy loss that support the
following qualitative conclusion:  The typical formation time $\tau$, needed for a parton to decohere and become 
an independent quanta, is set by the inverse of its transverse energy $\tau \sim \frac{\omega}{k_T^2}$
(measured with respect to the direction of its parent parton). In the dense QCD medium, the 
squared transverse momentum of the gluon grows by Brownian motion, $\left< k_T^2\right> \sim \hat{q}\, \tau$,
and the average formation time for partons of energy $\omega$ becomes
\be
\left< \tau \right> \sim \sqrt{\frac{\omega}{\hat q}} \, .
\ee 
So, in the QCD medium, softer gluons decohere from the parton shower at earlier time scales than harder ones
because medium-induced scattering decorrelates them faster from the direction of the parent parton.

The kinematical argument outlined above shows that soft jet components are easily transported out of a cone of 
finite radius. We now turn to the question whether these soft components can carry a sufficiently large fraction 
of the total jet energy outside the jet cone. Our discussion will be based solely on the mean energy 
distribution of partonic components inside a jet;  fluctuations around this mean remain to be considered in a
more detailed study. 

\begin{figure}
\centering
\includegraphics[angle=0,width=0.99\linewidth]{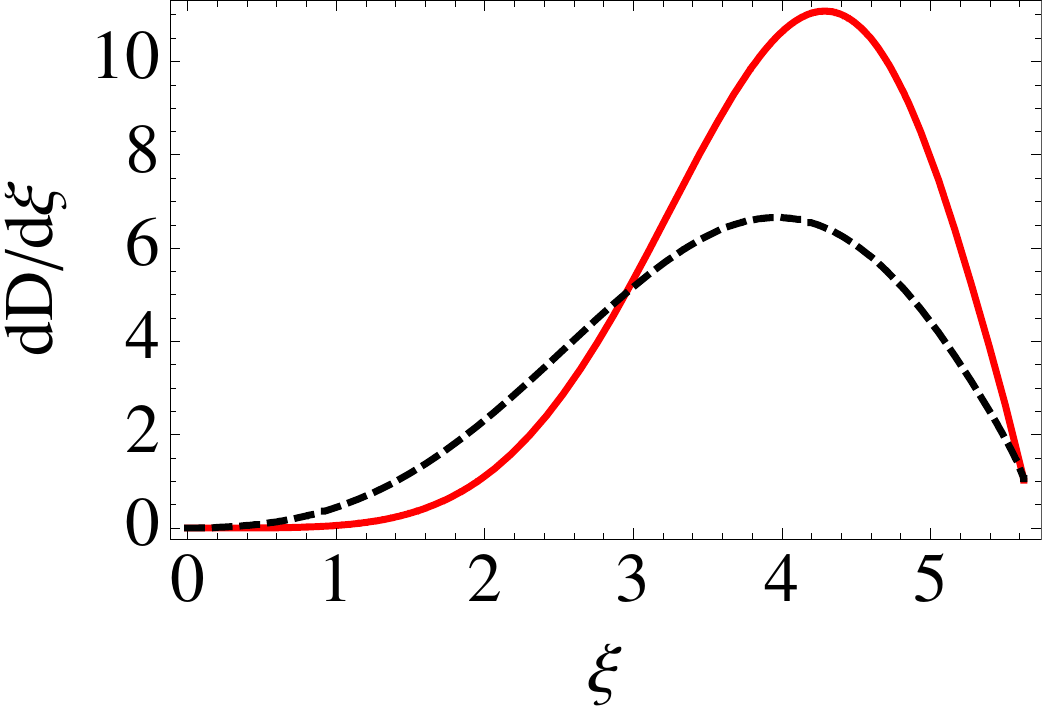}
\caption{
\label{fig:MLLA}
Dashed: Inclusive parton distribution in a jet of energy $E_T=100$ GeV obtained by MLLA evolution from an initial scale $Q_0=E_T \sin(R)$, ($R=0.4$) up to a final partonic scale $Q_f=1$ GeV. Solid: same distribution obtained by a medium modified MLLA kernel \cite{Borghini:2005em}. Note that due to kinematical constraints, there are no gluons with $z<E_T\sin\theta/Q$.
}
\end{figure}

\begin{figure}
\centering
\includegraphics[angle=0,width=0.99\linewidth]{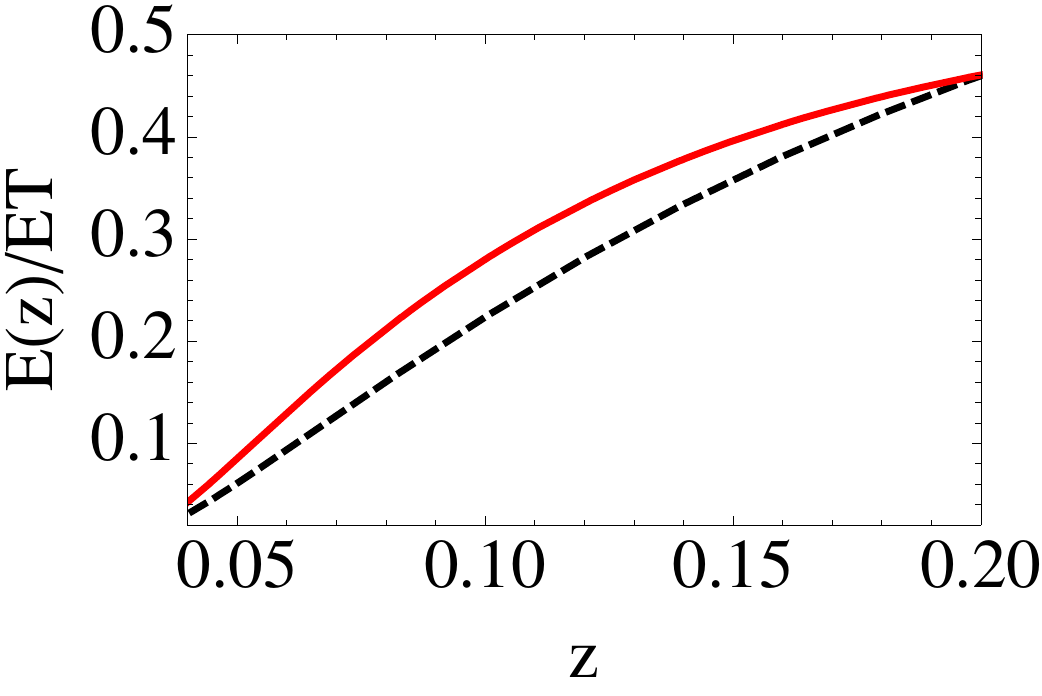}
\caption{
\label{fig:EMLLA}
Fraction of the total jet energy, $E_T$, carried by partons of energies less than $\omega=z E_T$ obtained from Fig. \ref{fig:MLLA} via Eq.~(\ref{eq:fracEdef}) in vacuum (dashed) and both in medium (solid). The plot does not extend to $z\rightarrow 1$ since MLLA is only valid at small z.
}
\end{figure}

For a parton shower evolving in the vacuum, intra-jet multiplicity distributions as a function of the longitudinal
momentum fraction $z \approx \omega/E_T$ can be calculated for sufficiently small values $z$ within the
modified leading logarithmic approximation (MLLA). These distributions are functions of both the jet energy
scale $E_T$ and the resolution scale $Q_f$ down to which components in the parton shower are counted. 
The single inclusive parton distribution as a function of $\xi=\log 1/z$  obtained by MLLA evolution
from an initial hard scale $Q_0=E_{T} \sin (R)$ down to a final scale $Q_f$ is shown in
Fig.~ \ref{fig:MLLA} (dashed line). Consistent with the praxis in Monte Carlo event generator
we choose $Q_f= 1$ GeV as the lowest scale for partonic evolution.  At this scale, one sees that
high energy jets contain already many soft partons.
From the single inclusive distribution, the average energy fraction of the jet carried by partons with energy 
fraction smaller than $z$ is given by
\be
\frac{E(z)}{E_T}=\int^\infty_{\log{1/z}} d\xi\, e^{-\xi} \frac{dD}{d\xi}\,,
\label{eq:fracEdef} 
\ee
which is shown in Fig. \ref{fig:EMLLA} (dashed line).
As long as the integration involves only soft components, $z \ll 1$, MLLA provides a good approximation
for $E(z)/ E_T$.

If the frequency collimation of soft partons is the sole medium modification, we can estimate $\hat{q} \, L$ by
determining from Fig. \ref {fig:EMLLA} the value $z$ for which the mean fractional energy coincides with the bounds 
on energy loss, Eq. (\ref{eq:lowerbound}) and Eq.~(\ref{eq:upperbound}). Since partons with energy
$\omega^2 = z^2\, E_T^2 \leq \hat q L$ are lost  from the cone, we obtain
\be
\label{q0est}
35 \,  \left(\frac{E_T}{E_0}\right)^2  \le\, \hat q L\, \le 85 \,  \left(\frac{E_T}{E_0}\right)^2  {\rm GeV^2}\,\, {\rm (estimate~1)}\, ,
\ee
with $E_T$ the jet energy given in units of $E_0 = 100$ GeV. 

A generic feature of all jet quenching models is the enhancement of small $z$ fragmentation partons as a consequence of medium induced gluon radiation. This effect is in particular necessary for 
describing the strong suppression of single inclusive hadron spectra measured at RHIC and the LHC~\cite{alice}.
Bound Eq.~(\ref{q0est}) neglects this effect, since it is based on a vacuum fragmentation function. 
By supplementing the MLLA framework with a medium-induced enhancement of parton branching, one 
can obtain simple models for the longitudinal softening of jet fragmentation functions~\cite{Borghini:2005em}.
An example of such a medium-enhancement which is roughly consistent with a factor 5 suppression of
leading hadron spectra, is shown in Fig. \ref{fig:MLLA} and Fig. \ref{fig:EMLLA} (solid lines).
Since there is a larger fraction of the total jet energy stored in soft components, a collimation up to the same
frequency $\sqrt{\hat q\, L}$ leads to a larger energy loss. In this case, we obtain
\be
\label{hqmmlla}
30 \,  \left(\frac{E_T}{E_0}\right)^2  \le\, \hat q L\, \le 60 \,  \left(\frac{E_T}{E_0}\right)^2  {\rm GeV^2}\, {\rm (estimate~2)}\, .
\ee

These estimates are subject to various uncertainties. Amongst the model-intrinsic ones, we mention  
the choice of the final resolution scale $Q_f$ that has 
significant impact on the distributions shown in Fig.~\ref{fig:MLLA} and Fig.~\ref{fig:EMLLA} (solid lines). 
Moreover, there are harder partonic components of order $\mathcal{O}
\left(\sqrt{\hat q L} / {R}\right)$ that will be
partially moved outside the jet cone. Including these components properly will require a discussion of fluctuations.  
Taking the above estimates at face value, an extraction of $\hat{q}$ demands information about the distribution
of in-medium path length. To arrive at first reference values, we note that 
$L\sim 6$ fm ($L\sim 10$ fm) yields $5 \le\ \hat q  \le 10 ~{\rm GeV^2/fm}$ ($3 \le\ \hat q  \le 6 ~{\rm GeV^2/fm}$).
It is clear, that these first estimates cannot replace detailed model studies that must account for both the suppression of
single inclusive hadron spectra \cite{alice} and the quenching of true jets~\cite{Atlas:2010bu,Bolek}.

{\bf 6. Conclusions.} In the discussion of RHIC data on single inclusive hadron suppression, emphasis was placed on
the strong longitudinal softening of the hardest parton in the shower. For this, radiative parton energy loss was identified
as the dominant mechanism. However, there was little experimental constraint so far on the angular distribution of
this radiation. As a consequence, models of radiative parton energy loss~\cite{Wiedemann:2009sh,CasalderreySolana:2007zz,d'Enterria:2009am,Majumder:2010qh,Jacobs:2004qv,Gyulassy:2003mc} did not focus on an accurate treatement of
transverse broadening. Nevertheless, some exploratory studies of jet broadening have been performed in the
framework of radiative energy loss~\cite{Salgado:2003rv,Lokhtin:2006dp,Vitev:2008rz}.

Here, we have stressed that trimming soft components away from the
jet parton shower via frequency collimation provides an efficient mechanism for reducing the
energy in the jet cone.  The main dynamical components invoked in this picture  have been implemented in various
models of jet quenching. In particular, collisional parton energy loss was identified in some model studies as a significant 
subleading mechanism~\cite{Wicks:2005gt}. Also, the analysis of medium responses to jet propagation 
led to the proposal of mechanisms that contribute to the transverse broadening of jet fragmentation in the 
medium~\cite{CasalderreySolana:2004qm,Li:2010ts}. 
Moreover, there has been a lot of activity in recent years on implementing these dynamical
mechanisms into Monte Carlo models of the final state parton shower~\cite{Zapp:2008gi,Zapp:2008af,Armesto:2009ab,Armesto:2009qg,Schenke:2009gb,Renk:2010zx,Wang:1991hta}, including the
role of decoherence and collisional broadening.
Finally, the importance of decoherence of the partonic shower, and of collisional mechanisms has also been
emphasized independently in~\cite{Leonidov:2010xf,MehtarTani:2010ma}.
In the present rapid communication, we have combined
these previously formulated key ideas in the dynamical picture of the medium as a
frequency collimator of jet parton showers. We have then provided semi-quantitative
estimates to support the view that frequency collimation
can account for the main features seen in the medium-induced dijet asymmetry
measurements by ATLAS. We expect that this observation can provide some guidance
for the further development of jet quenching models and Monte Carlo tools.

\noindent 
{\bf Acknowledgments.}
We thank Ilektra Christidi, Brian Cole, Peter Jacobs,
Tom LeCompte,  Andreas Morsch, J\"urgen Schukraft, Peter Skands and Bolek Wyslouch for useful discussions. 
One of us (JGM) acknowledges the support of Funda\c c\~ao para a Ci\^encia e a Tecnologia (Portugal) under project CERN/FP/109356/2009.

\end{document}